\documentclass{article}
\usepackage[utf8]{inputenc}
\usepackage{amsmath}
\usepackage{amsfonts}
\usepackage{braket}
\usepackage{graphicx}
\usepackage{physics}
\usepackage{xcolor}
\usepackage{multirow}
\usepackage{adjustbox}
\usepackage[a4paper, total={6in, 8in}]{geometry}

\title{Tunable giant two-dimensional optical beam shift from a tilted linear polarizer}
\author{Niladri Modak$^{1,*}$, Sayantan Das$^{1}$, Priyanuj Bordoloi$^{1}$, Nirmalya Ghosh$^{1,+}$\\
$^{1}$\textit{\small Department of Physical Sciences},\\ \textit{\small Indian Institute of Science Education and Research Kolkata,}\\ \textit{\small Mohanpur, India- 741246}}

\date{\footnotesize$^{*}$nm16ip018@iiserkol.ac.in, \footnotesize$^{+}$nghosh@iiserkol.ac.in}

\begin{document}

\maketitle
\begin{abstract}
  \noindent 
  We demonstrate an intriguing giant optical beam shift in a tilted polarizer system analogous to the Imbert-Fedorov shift in partial reflection around the Brewster's angle of incidence. We explain this giant shift using a generalized theoretical treatment for beam shift in a tilted uniaxial anisotropic material incorporating the three-dimensional orientation of its optic axis in the laboratory frame. We further demonstrate regulated control and tunability of both the magnitude and direction of the shifts by wisely tuning the input polarization and the corresponding orientation angles. These findings open up the possibility of precise micron-level beam steering using a simple linear polarizer.
  
\end{abstract}
The reflection and refraction of real optical beams at interfaces deviate from Snell's law when its cylindrical symmetry is broken \cite{bliokh2013goos}. In such a scenario, the beam, comprising of distribution of many wave vectors, exhibits polarization-dependent longitudinal and transverse deflections in the space domain (spatial shift) or in the momentum domain (angular shift) \cite{bliokh2013goos}. These so-called Goos–H\"anchen (GH) and Imbert–Federov (IF) beam shifts have evoked recent intensive investigations \cite{mazanov2021anomalous,wang2021shifting,bliokh2015quantum,bardon2019spin,lekenta2018tunable,wang2020probing}. 
While the GH shift originates from the angular gradient of the polarization-dependent reflection/refraction coefficients of the material, the IF shift has its origin in spin-orbit interaction of light appearing as a consequence of the evolution of momentum or space-gradient of the geometric phase leading to space or momentum domain beam shifts \cite{bliokh2013goos}. These beam shifts observed in diverse optical systems are not only fundamentally interesting \cite{bliokh2008coriolis,bardon2019spin,bliokh2016spin,takayama2018enhanced,zhu2021wave,mazanov2021anomalous,kim2021spin,gotte2014eigenpolarizations,toppel2013goos,kavokin2005optical,onoda2004hall,leyder2007observation,hosten2008observation,korger2014observation,aiello2009transverse} but also have  potential applications in metrology and have opened up novel route towards development of spin-orbit photonic devices \cite{su2020dynamically,wang2020probing,ling2017recent}. The magnitude and nature of the beam shifts in partial and total internal reflection strongly depend on the Fresnel reflection and transmission coefficients of the interface and on their angular dispersion \cite{bliokh2013goos}. In contrast, Aiello et al. introduced a novel variant of transverse beam shift, namely, geometrical spin hall effect of light (SHEL), which is independent of any material parameters and is purely geometric \cite{aiello2009transverse}. This was first experimentally demonstrated in a tilted linear polarizer system and has attracted particular attentions \cite{korger2014observation}. In a subsequent study, it was shown that such effect could be observed in a generalized tilted homogeneous uniaxial anisotropic system having either dichroism (polarizer) or retardance (waveplate) \cite{bliokh2016spin,bliokh2019spin}. Importantly, it was demonstrated that this beam shift is not purely geometric; rather, it depends upon the polarization anisotropy parameter of the sample. Understanding the nature of such intriguing beam shifts in tilted anisotropic systems, therefore, demands further investigations both on conceptual and practical grounds \cite{qin2010spin,takayama2018enhanced,bliokh2019spin,zhu2021wave}.
\par 
  In this paper, we experimentally demonstrate an intriguing giant optical beam shift in a tilted polarizer system analogous to the Imbert-Fedorov shift in partial reflection around the Brewster’s angle of incidence \cite{gotte2014eigenpolarizations}. 
  Complete 2D tunability of this giant beam shift is demonstrated further by tuning the input polarization state and the 3D orientation angles of the optic axis of the tilted polarizer.

 \par
\begin{figure}[h!]
\centering
\includegraphics[width=\linewidth]{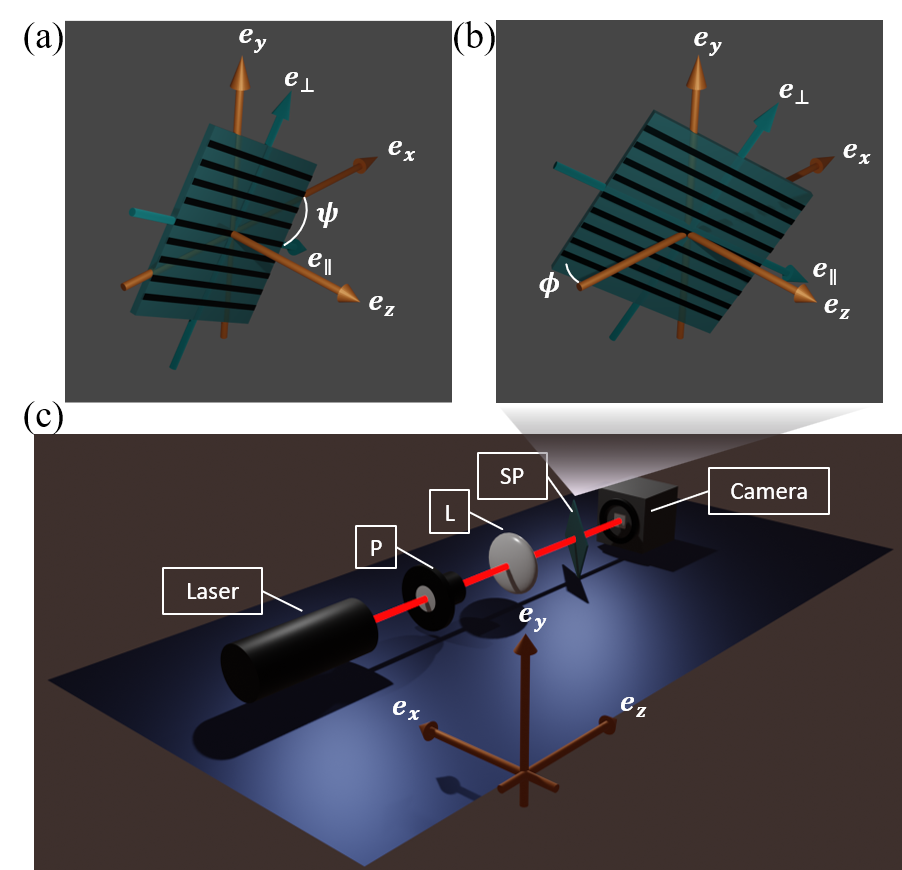}
\caption{Schematic illustration of the orientation of uniaxial anisotropic material. The anisotropic system ($\{\hat{e}_\parallel,\hat{e}_\perp\}$) is depicted as a teal blue plate. $\{x,y,z\}$ is the lab reference frame. Black lines on the plate indicate the direction of the optic axis. (a) The angle $\psi$ between $\hat{e}_x$ and $\hat{e}_{\parallel}$ describes azimuthal orientation of the optic axis. (b) $\phi$ is the tilt angle around $y$ axis. (c) Experimental setup. A 633 nm beam from a He-Ne laser source is passed through a Glan-Thompson linear polarizer P, a biconvex lens L (focal length 200 mm), and a sample sheet polarizer SP (LPVIS100, Thor Labs, USA) mounted in a tilted configuration ($\psi$ and $\phi$) and detected by a CCD camera (iKon M-912, pixel size= $24\ \mu m$, Andor, USA). The experimental parameters in Eq.\eqref{eq6} are noted in Sec.(S3) of SM.}
\label{fig1}
\end{figure}

To model the beam shifts in a general tilted uniaxial anisotropic system, we consider two sets of coordinate systems: one representing the lab frame $\{x,y,z\}$ with unit vectors $\{\hat{e}_x,\hat{e}_y,\hat{e}_z\}$ and the other describing the local reference frame $\{\parallel,\perp,\zeta\}$ of the uniaxial anisotropic material having unit vectors $\{\hat{e}_\parallel,\hat{e}_\perp,\hat{e}_\zeta\}$ (see Fig.\ref{fig1}(a) and (b)). Here, $\hat{e}_z$ is the direction of propagation of the input beam, and $\hat{e}_\parallel$ represents the direction of the optic axis of the uniaxial system. The azimuthal orientation $\psi$ of the optic axis in the local frame and the tilt angle $\phi$ around $y$ axis provide a complete description of the three-dimensional orientation of the optic axis of the uniaxial system in the lab frame (see Fig.\ref{fig1}(a) and (b)). Using this geometrical framework, the unit vectors $\hat{e}_{\parallel}$ and $\hat{e}_{\perp}$ in the lab frame can be obtained as (see Sec.(S1) of supplemental material (SM)).
\begin{equation}
    \hat{e}_{\parallel}=\begin{pmatrix} \cos{\psi}\cos{\phi}\\-\sin{\psi}\\\cos{\psi}\sin{\phi}\end{pmatrix};\ \hat{e}_{\perp}=\begin{pmatrix} \sin{\psi}\cos{\phi}\\\cos{\psi}\\\sin{\psi}\sin{\phi}\end{pmatrix} 
    \label{eq1}
\end{equation}
Accordingly, the angle between the direction of propagation of the beam and the optic axis becomes $\theta\equiv\cos^{-1}{(\hat{e}_{\parallel}\cdot \hat{e}_z})=\cos^{-1}{(\cos{\psi}\sin{\phi})}$. When a Gaussian beam passes through such a tilted anisotropic system with $\theta\neq0$, the output beam experiences polarization-dependent longitudinal and transverse shifts in the local frame \cite{bliokh2016spin} (see Sec.(s1) of the SM). The evolution of the electric field of the light beam through such a system can be modeled using the momentum domain Jones matrix $\boldsymbol{M}$ (under the first order paraxial approximation) \cite{bliokh2016spin}.
\begin{equation}
    \boldsymbol{M}=\begin{pmatrix} T_{\parallel}(1+\Theta_{\parallel} \mathfrak{X}_{\parallel}) & T_{\parallel}\Theta_{\perp} \mathfrak{Y}_{\parallel}\\
    -T_{\perp}\Theta_{\perp} \mathfrak{Y}_{\perp} & T_{\perp}(1+\Theta_{\parallel} \mathfrak{X}_{\perp})\end{pmatrix}
    \label{eq2}
\end{equation}
Here, $\mathfrak{X}_{\parallel\slash \perp}= \frac{d \ln{T_{\parallel\slash \perp}}}{d\theta}\  \text{and} \ \mathfrak{Y}_{\parallel\slash \perp}=(1-\frac{T_{\perp\slash \parallel}}{T_{\parallel\slash \perp}}) \cot{\theta}$. $T_{\parallel\slash \perp}$ represent complex amplitude transmission coefficients of the anisotropic material for the two orthogonal linear polarizations along $\hat{e}_{\parallel}/\hat{e}_{\perp}$ directions, and $\Theta_{\parallel\slash \perp}$ represents angular deviation or momentum spread of the beam along the respective directions \cite{bliokh2016spin}. In general, $T_{\parallel\slash \perp}$ encodes both amplitude anisotropy (dichroism for a polarizer) and phase anisotropy (retradnce for a waveplate) effects. As evident from Eq.\eqref{eq2}, $\boldsymbol{M}$ not only depends on the tilt angle $\phi$ but also on the local azimuthal orientation $\psi$ of the optic axis. For $\psi=0$, this general formalism reduces to the previously reported one \cite{bliokh2016spin,bliokh2019spin} (see Sec.(S1) of the SM). We shall subsequently illustrate the utility of using the two different angles in the context of controlling the magnitude and direction of beam shifts. Using the Jones matrix $\boldsymbol{M}$, polarization-dependent beam shifts along $\hat{e}_{\parallel}$ and $\hat{e}_{\perp}$ directions can be determined from the respective shift matrix operators as $\boldsymbol{G}$ $=\frac{i}{k}\begin{pmatrix} \mathfrak{X}_{\parallel} & 0 \\ 0 & \mathfrak{X}_{\perp}\end{pmatrix}$ and $\boldsymbol{S}$ $=\frac{i\cot{\theta}}{k}\begin{pmatrix} 0 & \mathfrak{Y}_{\perp}\\  -\mathfrak{Y}_{\parallel} & 0\end{pmatrix}$ (see Sec.(S1) of the SM) \cite{bliokh2016spin}. Here, $k$ is the magnitude of the central wave vector of the beam. The longitudinal shift (along $\hat{e}_{\parallel}$) appears due to the dispersion of the transmission coefficients with respect to the relative angle $\theta$ between the optic axis and the direction of propagation of the beam. The transverse shift (along $\hat{e}_{\perp}$), on the other hand, appears due to the evolution of geometric phase gradient which also depends on $\theta$ \cite{bliokh2016spin}. The physically observable beam shift for a given polarization $\ket{E}$ of the illuminating beam can be obtained using the expectation value of the corresponding shift operator $\mel{E_i}{\boldsymbol{G}/\boldsymbol{S}}{E_i}$ where $\ket{E_i}$ is input polarization. Note that here, the illuminating polarization $\ket{E}$ is modulated by the zeroth order Jones matrix $\boldsymbol{T}$ ($=\begin{pmatrix} T_{\parallel} &0 \\ 0 & T_{\perp} \end{pmatrix}$) (corresponding to the central wave vector $\vec{k}$) of the uniaxial anisotropic system as $\ket{E_i}=\boldsymbol{T}\ket{E}$. 

\par
We now turn to the specific case of a tilted polarizer, for which $T_{\parallel\slash \perp}$ are real. For an ideal polarizer, the angular dispersions of $T_{\parallel/\perp}$ are rather weak and accordingly the longitudinal shift is negligible \cite{bliokh2019spin,korger2014observation}. Henceforth, for the experimental perspective, we therefore consider contribution of the transverse shift operator $\boldsymbol{S}$ only. Thus, Eq.\eqref{eq2} for  a tilted polarizer with $\psi$ and $\phi$-orientation angles can be simplified to \cite{bliokh2019spin}
\begin{eqnarray}
    \boldsymbol{M}=(\mathbb{I}-ik\Theta_{\perp}\boldsymbol{S})\boldsymbol{T}=(\mathbb{I}-ik[\Theta_x\boldsymbol{S}_x+\Theta_y\boldsymbol{S}_y+\Theta_z\boldsymbol{S}_z&])\boldsymbol{T} \nonumber\\
    \text{with}\ \boldsymbol{S}_x=\sin{\psi}\cos{\phi}\boldsymbol{S};\boldsymbol{S}_y=\cos{\psi}\boldsymbol{S};\boldsymbol{S}_z=\sin{\psi}\sin{\phi}\boldsymbol{S}& \label{eq3}
\end{eqnarray}
Clearly, the observable beam shifts along the lab $x,\ y,\ \text{and}\ z$ directions are determined by the $\psi$ and $\phi$-dependent components of the shift matrix, $\boldsymbol{S}_x$, $\boldsymbol{S}_y$, and $\boldsymbol{S}_z$ respectively. The magnitude of these shifts can be determined from the eigenvalues $d_{\pm}$ of $\boldsymbol{S}$ with corresponding eigenpolarizations $\ket{r_\pm}$.
\begin{eqnarray}
    d_{\pm}&=&\mp\frac{i \cot{\theta}}{k}(\sqrt{T_{\parallel}/T_{\perp}}-\sqrt{T_{\perp}/T_{\parallel}})\nonumber\\ \ket{r_\pm}&=&\frac{1}{\sqrt{T_{\parallel}+T_{\perp}}}\begin{pmatrix}\sqrt{T_{\parallel}}\\ \pm \sqrt{T_{\perp}} \end{pmatrix}
    \label{eq4}
\end{eqnarray}
As usual, the imaginary eigenvalues will be manifested as angular shift of the output beam \cite{gotte2014eigenpolarizations} and the corresponding eigenpolarizations are linear here (as $T_{\parallel},\ T_{\perp}$ are real for a polarizer). Importantly, the eigenvalues $d_{\pm}$ will become exceedingly large for an ideal polarizer ($T_{\perp}/T_{\parallel}\gg1$). This very feature of a polarizer enables the beam to experience a giant deflection $D=d_+ - d_-$ along local transverse direction $\hat{e}_{\perp}$ when the incident polarizations match with the eigenpolarizations of $\boldsymbol{S}$ (see Sec.(S2) of the SM).
\par
The above scenario of giant beam shift for incident eigenpolarization is similar in spirit to that observed for light beam partially reflected from an interface around the Brewster's angle of incidence \cite{gotte2014eigenpolarizations}. In general, the transverse shift matrix operator in case of partial reflection can be written as a complex combination of the Pauli matrices $\sigma_2$ and $\sigma_1$ \cite{Jayaswal:14,bliokh2013goos}. While $\sigma_2$ represents SHEL shift with circular eigenpolarizations, $\sigma_1$ describes the IF shift with diagonal linear polarizations as eigenstates \cite{gupta2015wave,bliokh2013goos}. As one approaches the Brewster's angle of incidence from above ($>$ Brewster's angle), the eigenvalues of the transverse shift operator remain real yielding space domain beam shift and the corresponding eigenpolarizations are extreme elliptical. Conversely, when Brewster's angle is approached from below ($<$ Brewster's angle), the eigenvalues become imaginary (yielding angular shift) and the corresponding eigenpolarizations become linear. Our situation is thus in close analogy to the latter scenario. The transverse shift here is angular in nature with linear polarizations as eigen states \cite{gotte2014eigenpolarizations}.
\par 
Following Eq.\eqref{eq3}, the components of $D$ in the lab frame $\{x,y,z\}$ become
\begin{equation}
    D_x=\sin{\psi}\cos{\phi}D;D_y=\cos{\psi}D;D=\sin{\psi}\sin{\phi}D
    \label{eq5}
\end{equation}
The angular shifts $\Theta_{x,y,z}$ (corresponding to imaginary $D$) in the lab frame between two eigenpolarizations $\ket{r_\pm}$ can be obtained from Eq.\eqref{eq5}. The corresponding physically observable eigenshifts $\Delta_{x,y,z}$ of the beam centroid is related as
\begin{equation}
    \Theta_{x,y,z}=\frac{1}{Z_0}\Im(D_{x,y,z});\Delta_{x,y,z}=Z\Theta_{x,y,z}
    \label{eq6}
\end{equation}
Here, $Z$ is the propagation distance of the beam after passing through the tilted polarizer and $Z_0$ is its Rayleigh range. As evident from Eq.\eqref{eq6}, the locally transverse $D$ imparts its counterparts in the lab $x$, $y$, and $z$-direction, with their magnitudes determined by the $\psi$ and $\phi$ dependent pre-factors. Using our experimental embodiment with a 2D detector (see Fig.\ref{fig1}(c)), $\Delta_x$ and $\Delta_y$ can only be detected. In what follows, we therefore proceed to experimentally demonstrate large eigenpolarization beam shifts $\Delta_x$ and $\Delta_y$ and their tunability by suitably controlling $\psi$ and $\phi$.
\par
\begin{figure}[h!]
\centering
\includegraphics[width=\linewidth]{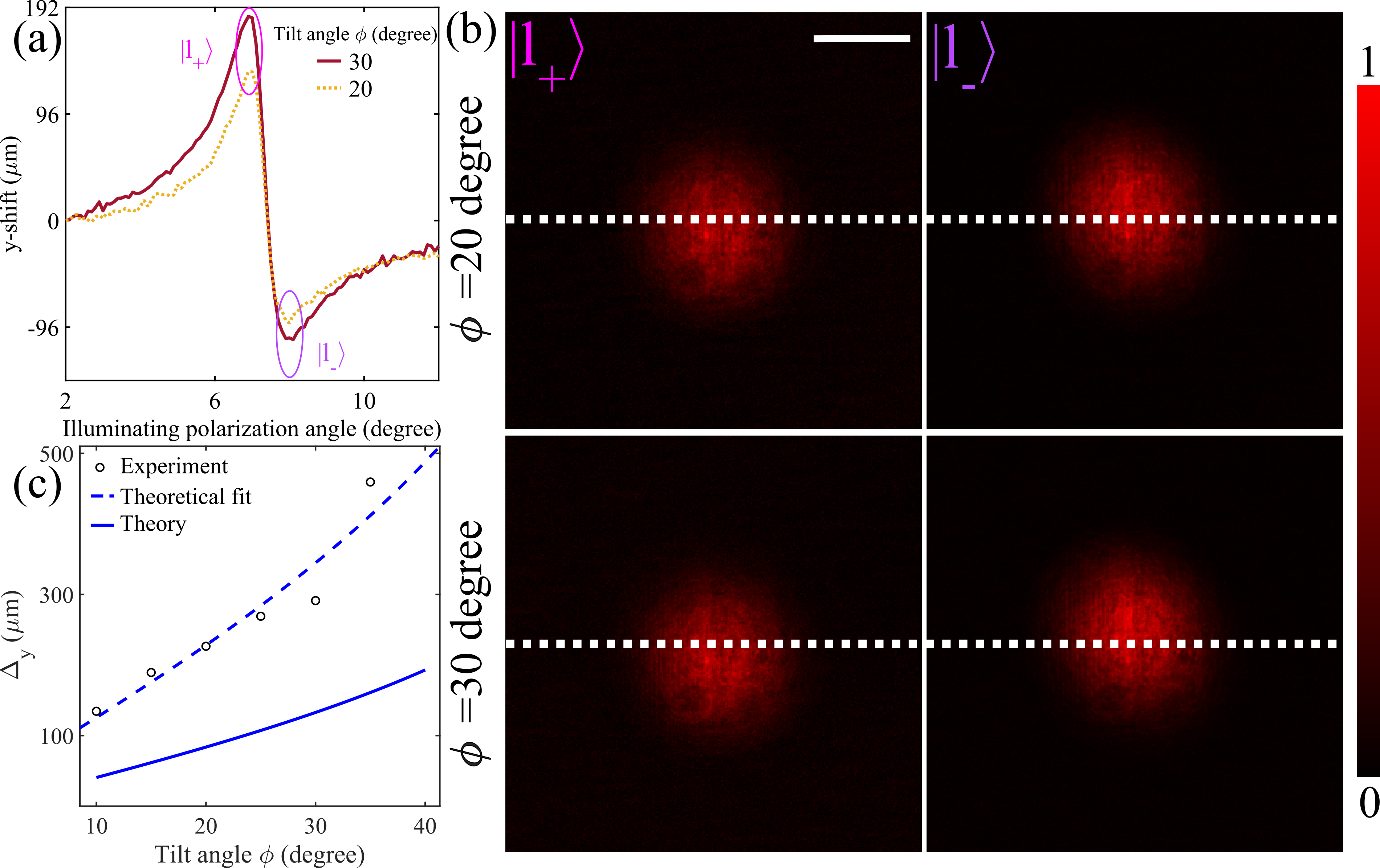}
\caption{Dependence of the experimental transverse ($y$) shift on the tilt angle $\phi$ for fixed azimuthal orientation of the optic axis of the polarizer at $\psi=0^{o}$. (a) Variation of the experimentally observed $y$-shift with varying angle of polarization of the illuminating beam for two representative tilt angles, $\phi=20^{o}$ (yellow dashed line) and $\phi=30^{o}$ (brown solid line). The angles of illuminating polarization ($6.9^{o},8.0^{o}$ for $\phi=20^{o}$ and $6.9^{o},8.1^{o}$ for $\phi=30^{o}$) corresponding to the left eigenpolarizations $\ket{l_{+,-}}$ (for eigenshift $\Delta_y$) are marked by magenta, violet ellipses inside the figure. The noted illuminating polarization angle are not the absolute angle of polarization in the lab frame, but are the angle of the rotational mount attached with preselecting polarizer P. (b) The intensity normalized beam profiles generating 
$\ket{l_{+}}$ and $\ket{l_{-}}$ (left and right panel) for $\phi=20^{o}$ (top panel) and $\phi=30^{o}$ (bottom panel). White scale-bar: $1200\mu m$. (c)The variation of $\Delta_y$ as a function of $\phi$. Black circle: experimental data. Blue dashed line: corresponding theoretical fit with Eq.\eqref{eq6}. The corresponding theoretical predictions of the exact magnitudes of $\Delta_y$ are also shown additionally (solid line).}
\label{fig2}
\end{figure}

First, we investigate the shifts by fixing the azimuthal orientation of SP at $\psi=0^{o}$ and varying the tilt angle $\phi$. Here, $\psi=0^{o}$ essentially implies that only $\Delta_y\neq0$. The polarization of the illuminating beam $\ket{E}$ is selected and controlled by P, which needs to be set at the left eigenpolarizations $\ket{l_\pm}$ of $\boldsymbol{S}$ to observe the giant eigenshift (see Sec.(S2) of the SM). The results of the variation of the $y$-shift of the beam centroid are summarized in Fig.\ref{fig2}. $\ket{l_\pm}$ correspond to the angle of the input linear polarization for which the $y$-shift of the beam centroid is observed to be maximum, and are accordingly marked in the figure for two tilt angles, $\phi=20^{o}$ and $30^{o}$. Note, the difference of $y$-shift between the two extreme points provides $\Delta_y$.   
The observed large $\Delta_y$ of the beam centriod for the eigenpolarization states are even visually perceptible from the beam profiles across the white dotted reference lines for both $\phi=20^{o},\ \text{and}\ 30^{o}$ (see Fig.\ref{fig2}(b)). Fig.\ref{fig2}(c) illustrates the dependence of the corresponding eigenshift $\Delta_y$ on $\phi$. The experimental data are fitted with $\Delta_y=a \tan{\phi}+b$ (obtained using Eq.\eqref{eq6}). In general, the shifts are large and the fitted value $a\approx536\mu m$ appears to be in agreement to the exact theoretical prediction $a\approx230\mu m$ (see Sec.(S3) of the SM).
\par
The above results provide experimental evidence of the giant beam shift for eigenpolarization of light beam passing through a tilted polarizer. We emphasize that this giant beam shift is not a manifestation of weak value amplification \cite{aharonov1988result,duck1989sense,ritchie1991realization,dressel2014colloquium,kofman2012nonperturbative,hosten2008observation} and there is no additional post-selection involved in our experiment unlike in \cite{bliokh2019spin}. Moreover, as evident from the experimental parameters, we are not even in the weak coupling regime which requires the difference between the eigenvalues to be much less than the spread of the pointer \cite{aharonov1988result,duck1989sense,dressel2014colloquium,kofman2012nonperturbative,hosten2008observation}. On the contrary, here the magnitudes of the eigenshifts are several times larger than the FWHM of the Gaussian beam or the beam waist incident on the SP ($\approx$ 20 $\mu m$). Note that similar to the Brewster's scenario in partial reflection \cite{gotte2014eigenpolarizations}, here also, the eigenshifts appear around the region of minimum intensity of the transmitted beam (see Sec.(S2) of the SM).


\begin{figure}[h!]
\centering
\includegraphics[width=\linewidth]{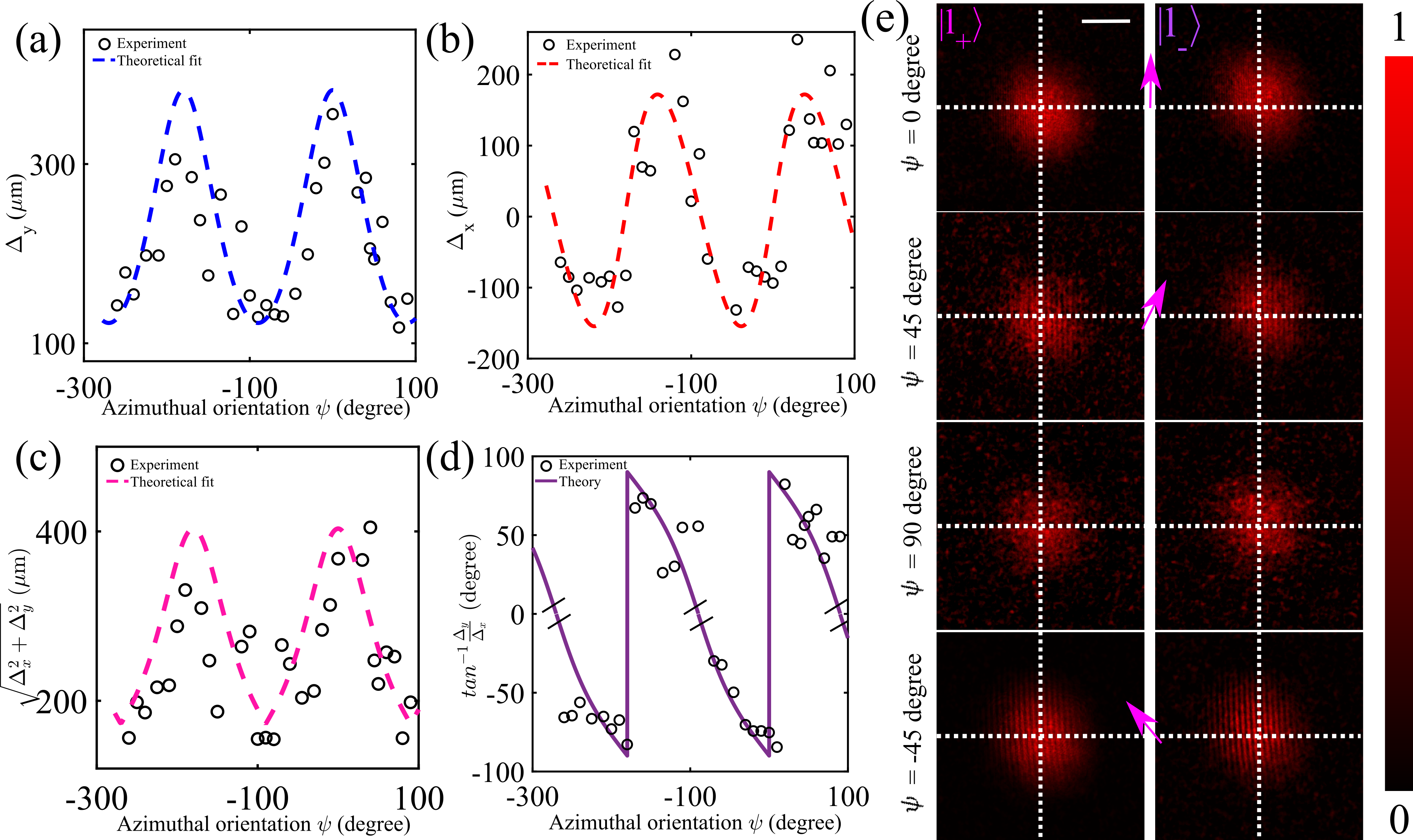}                                   
\caption{Dependance of eigenshifts $\Delta_y,\Delta_x$ on the azimuthal orientation $\psi$ of the optic axis of the polarizer tilted at $\phi=50^{o}$. (a) Variation of $\Delta_y$ with changing $\psi$. Black circles: experimental data. Blue dashed line: theoretical fit with Eq.\eqref{eq6}. (b) Variation of $\Delta_x$ with changing $\psi$. Black circles: experimental data. Red dashed line: corresponding theoretical fit with Eq.\eqref{eq6}. (c) The dependence of net beam shift $\sqrt{\Delta_x^2+\Delta_y^2}$ on $\psi$. Black circles: experimental data, magenta dashed line: theoretical fit. (d) The direction of the shift $\tan^{-1} ({\Delta_y/\Delta_x})$ on the $xy$ plane is plotted with changing $\psi$. Black circles: experimental data; violet solid line: theoretical prediction. (e) The intensity normalized beam profiles generating two eigenpolarizations $\ket{l_{+}}$ and $\ket{l_{-}}$ (left and right panel) for four different $\psi$ (different rows). White scale bar: $1200\mu m$. Magenta arrows indicate the corresponding directions of shifts.}
\label{fig3}
\end{figure}


\par
We now proceed to study the influence of the azimuthal orientation $\psi$ of the tilted polarizer on the beam shift. It is evident from Eq.\eqref{eq6}, for $\psi\neq0^{o}$, one will have contributions of both $\Delta_x$, $\Delta_y$. We keep $\phi=50^{o}$ and vary $\psi$ over its entire range to obtain full range of variations of $\Delta_x$ and $\Delta_y$ (see Fig.\ref{fig3}). 
Several interesting trends can be gleaned from the observed dependence of $\Delta_x$, $\Delta_y$ on $\psi$. $\Delta_y$ attains its maximum around $\psi=n\pi;n=0,\pm1,\pm2,\ldots$,  (see Fig.\ref{fig3}(a)). The magnitudes of the shift are also found to be in good agreement with the corresponding theoretical predictions. 
$\Delta_x$, on the other hand, is observed to be much weaker as compared to $\Delta_y$, and it assumes both `$+$'ve and `$-$’ve values as anticipated from Eq.\eqref{eq6} (see Fig.\ref{fig3}(b)). Relatively less agreement with the theory in this case possibly arises due to the much smaller magnitude of $\Delta_x$ (corresponds to only $\sim3-4$ pixels of the CCD camera). 
The maximum magnitude of the net beam shift $\sqrt{\Delta_x^2+\Delta_y^2}$ in the $xy$ plane is obtained at $\psi=n\pi;n=0,\pm1,\pm2,\ldots$ (see Fig.\ref{fig3}(c)). 
With regard to the direction of the observed shift $\tan^{-1} {\Delta_y/\Delta_x}$ (see Fig.\ref{fig3}(d)), it is noted that around $\psi=(2n+1)\pi/2;\ n=0,\pm1,\pm2,\ldots$, both $\Delta_x,\Delta_y\rightarrow0$ and hence the direction of the shift $\tan^{-1} {\Delta_y/\Delta_x}$ is undefined (marked with black parallel lines in Fig.\ref{fig3}(d)). Accurate experimental detection of the shifts and obtaining their directions around these regions are very challenging. On the other hand, the shifts are quite prominent around $\psi=(2n+1)\pi/2;\ n=0,\pm1,\pm2,\ldots$. The observed discontinuity in $\tan^{-1} {\Delta_y/\Delta_x}$ at $\psi=n\pi;n=0\pm1,\pm2,\ldots$ arises due to the choice of the range of principle values of the $\tan^{-1}$ function to lie within the interval ($-\pi/2,\pi/2$). These experimental results confirm that the introduction of the local azimuthal orientation of the tilted polarizer modifies the magnitude and the direction of locally transverse beam shifts in the lab frame. This offers regulated 2D control of the beam shift. 
\par
\begin{figure}[h!]
\centering
\includegraphics[width=\linewidth]{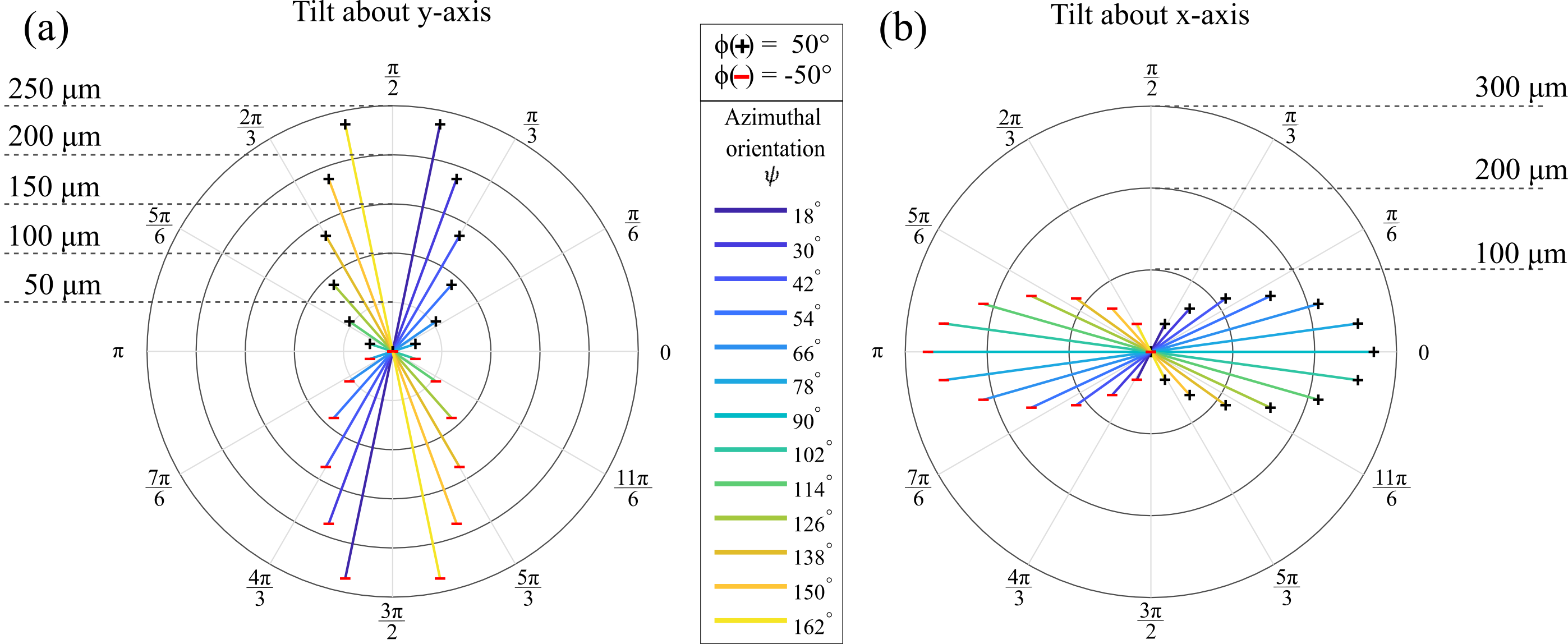}
\caption{Polar illustration of the amplitude and direction of the net eigenshifts in $xy$ plane from a tilted polarizer for varying azimuthal orientation $\psi$ and tilt angle $\phi$. `$+$'-headed arrows: the shift for $\phi=+50^{o}$, `$-$'-headed arrows: the shift for $\phi=-50^{o}$. Length of the arrows: the magnitude of shifts, polar angle of the arrows: the direction of the shifts. Different coloured arrows indicate eigenshifts for different values of $\psi$. Direction and magnitude of eigenshifts for the tilt about (a) $y$ axis and (b) $x$ axis.}
\label{fig4}
\end{figure}

  It is however, apparent from Fig.\ref{fig3}(c) and (d) that by regulating the angles $\psi$, $\phi$, and the input polarization state, one can only obtain a limited tuning of the beam shift on the lab $xy$ plane. The beam shift along $x$ direction in the lab frame is not achievable by imparting the tilt around $y$ axis. This limitation can be overcome if we introduce the tilt $\phi$ around $x$ axis separately as we demonstrate now. 
  From Fig.\ref{fig4}, it is evident that the tilt around $\hat{y}$ (Fig.\ref{fig4}(a)) complements the tilt around $\hat{x}$ (Fig.\ref{fig4}(b)) (see Sec.(S1) of the SM) for a fixed value of $\phi$. These results clearly demonstrate that by introducing both the tilts and by judiciously selecting the illuminating polarization state $\ket{E}$, in this simple experimental embodiment, one can obtain completely tunable giant beam shifts in the transverse $xy$ plane.
  

In summary, we have demonstrated an intriguing giant optical beam shift in a tilted linear polarizer system. We have understood this beam shift using a robust theoretical framework for optical beam shift in a tilted anisotropic system that encompasses general 3D orientation of the optic axis. Remarkably, the experimental beam shifts are observed to be $\sim300\mu m$, almost $10$ times the beam waist of the interacting Gaussian beam. This giant beam shift is interpreted as an eignepolarization shift of the corresponding shift operator for an ideal tilted linear polarizer, and an interesting analogy with the transverse beam shift for partial reflection at Brewster's angle of incidence is found. We have experimentally demonstrated regulated 2D control over such beam shift by wisely tuning the input polarization and the two different angles describing the 3D orientation of the optic axis of the tilted polarizer. The attainment of complete tunability of 2D giant beam shift using a linear polarizer in a simple experimental embodiment may have potential applications in precision beam steering \cite{salazar2015demonstration}.

\section*{Supplemental material}

\subsection*{S1. Theoretical framework for polarization dependent tunable beam shifts from a uniaxial anisotropic material}
\label{s1}
\begin{figure}[h!]
\centering
\includegraphics[width=0.3\linewidth]{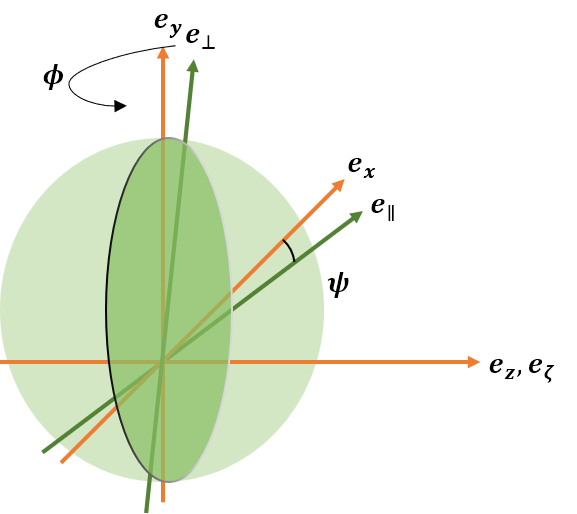}
\caption{Schematic illustration of the rotation introduced to evolve the optic axis in three dimension when the tilt is given around $y$ axis. $\{x,y,z\}$ is the laboratory reference frame, $\{\parallel,\perp,\zeta\}$ is the local frame of the material.}
\label{figs1aa}
\end{figure}

 \paragraph{a.}We have represented the local co-ordinates $\hat{e}_{\parallel}$ and $\hat{e}_{\perp}$ of the anisotropic materials in the laboratory frame $\hat{e}_{x}$, $\hat{e}_{y}$, and $\hat{e}_{z}$ in the main text. Now we demonstrate that how $\hat{e}_{\parallel}$ and $\hat{e}_{\perp}$  in the laboratory frame can be obtained by consecutive operation of two rotation matrices on $\hat{e}_{x}$ and $\hat{e}_{y}$ respectively (see Fig.\ref{figs1aa}): rotation $-\psi$ around $\zeta$ axis $R_\zeta(-\psi)$ and rotation $-\phi$ around $y$ axis $R_y(-\phi)$. 
\begin{equation}
\hat{e}_{\parallel/\perp}=R_y(-\phi)R_\mathfrak{z}(-\psi) \hat{e}_{x/y}
\label{seq1}
\end{equation}

The optic axis swipes the full area of the deep green disk under a full evolution of $\psi\ \text{from}\ 0\ \text{to}\ 2\pi$ (see Fig.\ref{figs1aa}). This disk swipes full the volume of the light green sphere when $\phi$ goes from $0$ to $2\pi$. Thus one can locate the optic axis (along $\hat{e}_{\parallel}$) in full three dimension using two consecutive rotation $R_\zeta(-\psi)$ and $R_y(-\phi)$.
\paragraph{b.}We have introduced the azimuthal orientation $\psi$ in addition to the previously reported tilt angle $\phi$ \cite{korger2014observation,bliokh2016spin,bliokh2019spin}. As Bliokh et al. predicted, the longitudinal and transverse shift appears when there is a non-zero angle $\theta$ between optic axis of the system and direction of propagation of the central wave vector $\vec{k}$ of the beam \cite{bliokh2016spin,bliokh2019spin}. In our case, $\theta$ depends on both $\psi$ and $\phi$. At $\psi=0$, $\theta$ becomes $\pi/2-\phi$ which was obtained previously by Bliokh et al. \cite{bliokh2016spin,bliokh2019spin}. From Eq. (2) of the main text, $\boldsymbol{M}$ suggests polarization dependent locally longitudinal (along $\hat{e}_{\parallel}$) and transverse (along $\hat{e}_{\perp}$) shift.
\begin{equation}
    \boldsymbol{M}=(\mathbb{I}-ik[\Theta_{\parallel}\boldsymbol{G}+\Theta_{\perp}\boldsymbol{S}])\boldsymbol{T}
    \label{seq2}
\end{equation}
Where $\boldsymbol{G}$ is the shift matrix for longitudinal shift, $\boldsymbol{S}$ is that for transverse shift, $\boldsymbol{T}$ is the zeroth order Jones matrix of the anisotropic system under consideration. $k$ is the central wavevector of the incident beam. As these local coordinate does not coincide with the loboratory $\{x,y,z\}$ frame, both the local shifts have their components along $\hat{e}_x$, $\hat{e}_y$, and $\hat{e}_z$.  Both the angles  $\psi$ and $\phi$ control the magnitude and direction of the shifts in the global frame. For the example of the tilted polarizer, the locally transverse shift is manifested in the laboratory frame depending on the value of $\psi$ and $\phi$ (see Eq.(5) and (6) of the main text).

\begin{figure}[h!]
\centering
\includegraphics[width=0.5\linewidth]{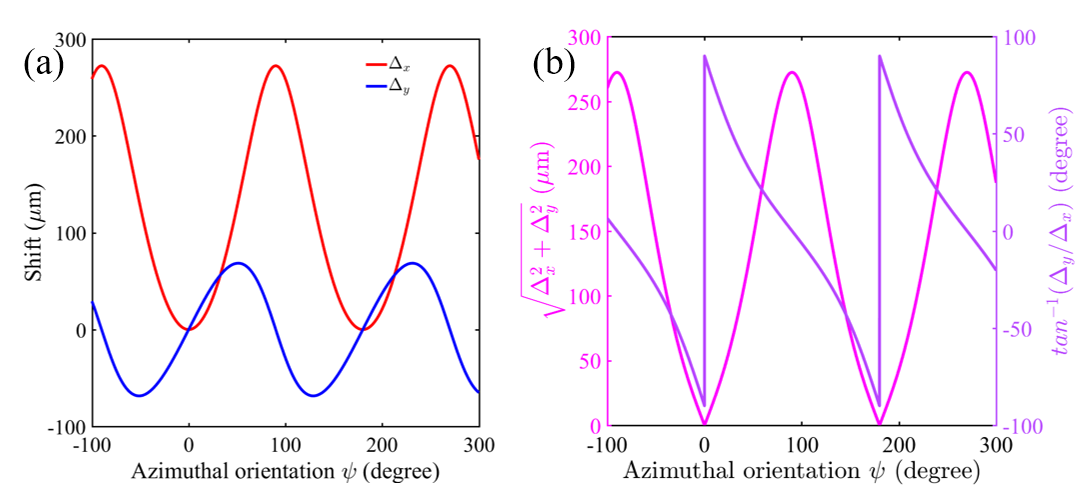}
\caption{The theoretically obtained variation of the eigen shifts with changing azimuthal orientation $\psi$ at $\phi=50^{o}$ when the tilt is imparted around $x$ axis. (a) Variation of $\Delta_x$ (red solid line) and $\Delta_y$ (blue solid line) with changing $\psi$. (b)  The dependence of resultant magnitude of the shift $\sqrt{\Delta_x^2+\Delta_y^2}$ (magenta solid line) and the direction of the shift $\tan^{-1} {\Delta_y/\Delta_x}$ (violet solid line) on $\psi$. These plots are obtained from Eq.\eqref{seq5}. The parameters are taken same as mentioned in Sec.(S3)}.
\label{figs1a}
\end{figure}
\paragraph{c.}As noted in the main text, tilt $\phi$ around $y$-axis provides only limited tunability of the beam shift in the laboratory frame which can be overcome by imparting the tilt $\phi$ around $x$-axis. Now we reiterate the framework for the tilt around $x$-axis. In that case, the unit vectors of the local reference frame of the system can be written as follows. 
\begin{equation}
\hat{e}_{\parallel/\perp}=R_x(-\phi)R_\mathfrak{z}(-\psi) \hat{e}_{x/y}
\label{seq3}
\end{equation}
The unit vectors take the following form in the laboratory frame.
\begin{equation}
    \hat{e}_{\parallel}=\begin{pmatrix} \cos{\psi}\\-\cos{\phi}\sin{\psi}\\\sin{\phi}\sin{\psi}\end{pmatrix};\ \hat{e}_{\perp}=\begin{pmatrix} \sin{\psi}\\\cos{\phi}\cos{\psi}\\-\cos{\psi}\sin{\phi}\end{pmatrix} 
    \label{seq4}
\end{equation}
Using Eq.\eqref{seq4} and following Eq.(6) of the main text, the eigenshifts in the laboratory frame can be written as
\begin{equation}
    \Delta_{x,y,z}=\frac{Z}{Z_0}\Im(D_{x,y,z})
    \label{seq5}
\end{equation}

\begin{figure}[h!]
\centering
\includegraphics[width=0.7\linewidth]{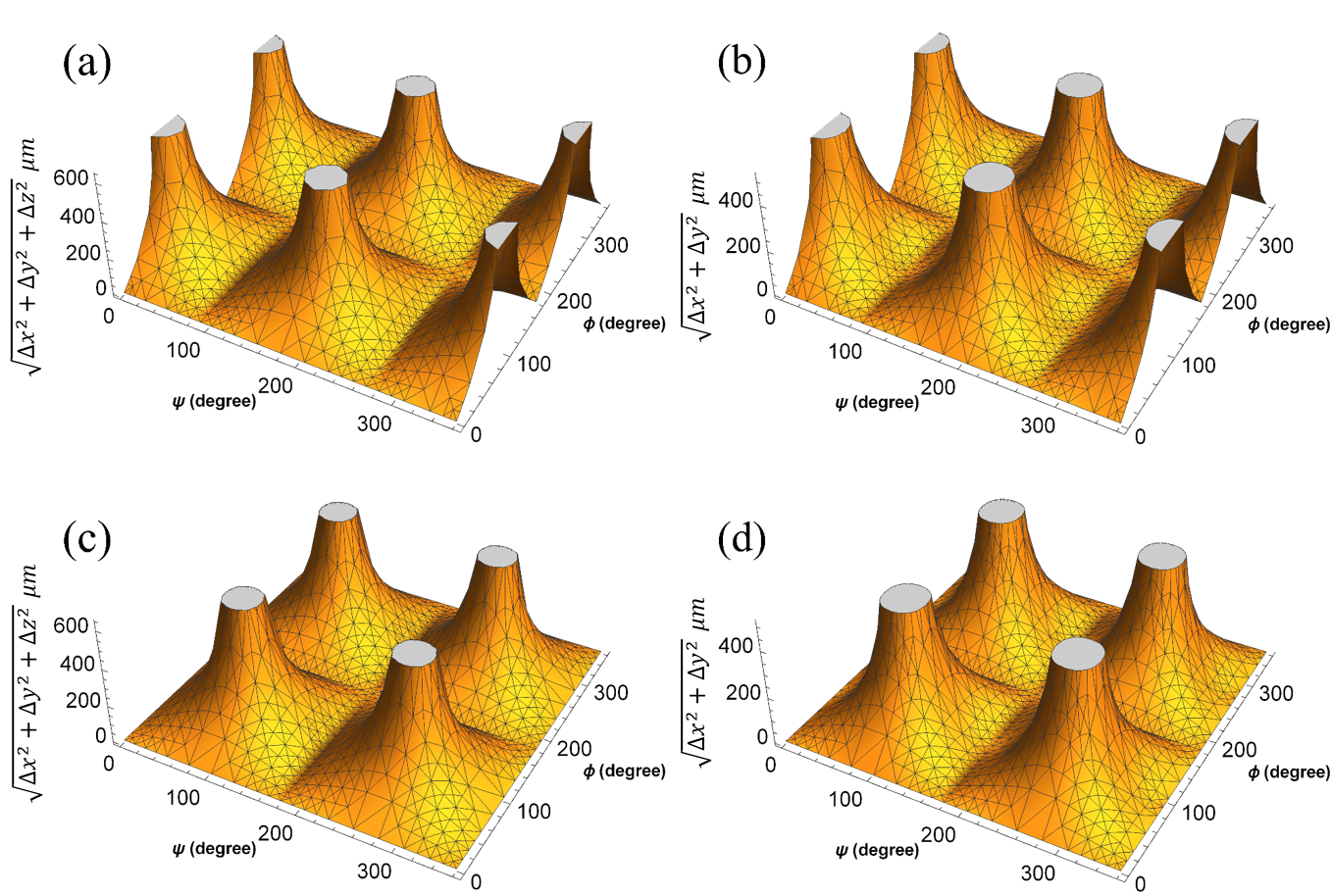}
\caption{The theoretically obtained variation of the magnitude of total eigen shifts $\sqrt{\Delta_x^2+\Delta_y^2+\Delta_z^2}$ and the magnitude of eigen shifts in $xy$ plane $\sqrt{\Delta_x^2+\Delta_y^2}$ with changing $\psi$ and $\phi$ for tilt around $y$ and $x$ axis. (a) Variation of $\sqrt{\Delta_x^2+\Delta_y^2+\Delta_z^2}$ when the tilt is given around $y$ axis. (b)  The dependence of $\sqrt{\Delta_x^2+\Delta_y^2}$ on $\psi$ and $\phi$ when the tilt is given around $y$ axis. (c), and (d) Variation of the $\sqrt{\Delta_x^2+\Delta_y^2+\Delta_z^2}$ and $\sqrt{\Delta_x^2+\Delta_y^2}$ with changing $\psi$ and $\phi$ when the tilt is given around $x$ axis. The divergence in the magnitude of the shifts can be overcome by including higher order terms of the shift matrices \cite{gotte2013limits}. The eigenshifts get maximized and minimized at the complementary values of $\psi$ and $\phi$ for the tilt around $y$ and $x$ axis. The parameters are taken same as mentioned in Sec.(S3).}
\label{figs3}
\end{figure}
Fig. \ref{figs1a}(a) demonstrates the variation of eigen shifts $\Delta_x$ (red solid line) and $\Delta_y$ (blue solid line) with changing the azimuthal orientation $\psi$ at a fixed tilt $\phi=50^{o}$ around $x$ axis. In Fig.\ref{figs1a}(a), the magnitude of resultant shift $\sqrt{\Delta_x^2+\Delta_y^2}$ is plotted in magenta solid line with changing $\psi$. The direction of the shift $\tan^{-1} ({\Delta_y/\Delta_x})$ is plotted in violet solid line. Comparing Fig.\ref{figs1a} and Fig.3 of the main text, one can conclude that the tilt around $x$ axis and $y$ axis are complementary to each other by means of the direction of the eigenshifts. 
\par 
We also plot the magnitude of the resultant shift in three dimension $\sqrt{\Delta_x^2+\Delta_y^2+\Delta_z^2}$ and that in the $xy$ plane $\sqrt{\Delta_x^2+\Delta_y^2}$ with changing $\psi$ and $\phi$ for the tilt around $y$ (see Fig.\ref{figs3}(a) and (b)) and $x$ (see Fig.\ref{figs3}(c) and (d)) axis. Fig.\ref{figs3} is an evidence of complementarity of the magnitude of the eigen shifts for the tilt around $y$ and $x$ axis. In conclusion, wisely tuning both these tilts and selecting the input polarizations, one can steer the beam in $xy$ plane.

\subsection*{S2. Connection of the polarization of the incident beam to the eigen polarizations of $\boldsymbol{S}$}
\label{s2}
$\boldsymbol{S}$ is a non-Hermitian matrix and hence, one can distinguish between its left and right eigen polarizations. The right eigen polarizations are given in Eq.(4) of the main text. The corresponding left eigen polarizations are
\begin{equation}
    \ket{l_\pm}=\frac{1}{\sqrt{T_{\parallel}+T_{\perp}}}\begin{pmatrix}\sqrt{T_{\perp}}\\ \pm \sqrt{T_{\parallel}} \end{pmatrix}
    \label{eqs1}
\end{equation}
When the illuminating polarizer P (see Fig.1 of the main text) matches with $\ket{l_\pm}$, then the zeroth order Jones matrix $\boldsymbol{T}$ of the tilted polarizer acts on it to modify it into $\ket{r_\pm}$ which then interacts with the shift matrix. Hence one can differentiate between illuminating polarization $\ket{E}$ and input polarization $\ket{E_i}$. 
\begin{figure}[h!]
\centering
\includegraphics[width=0.3\linewidth]{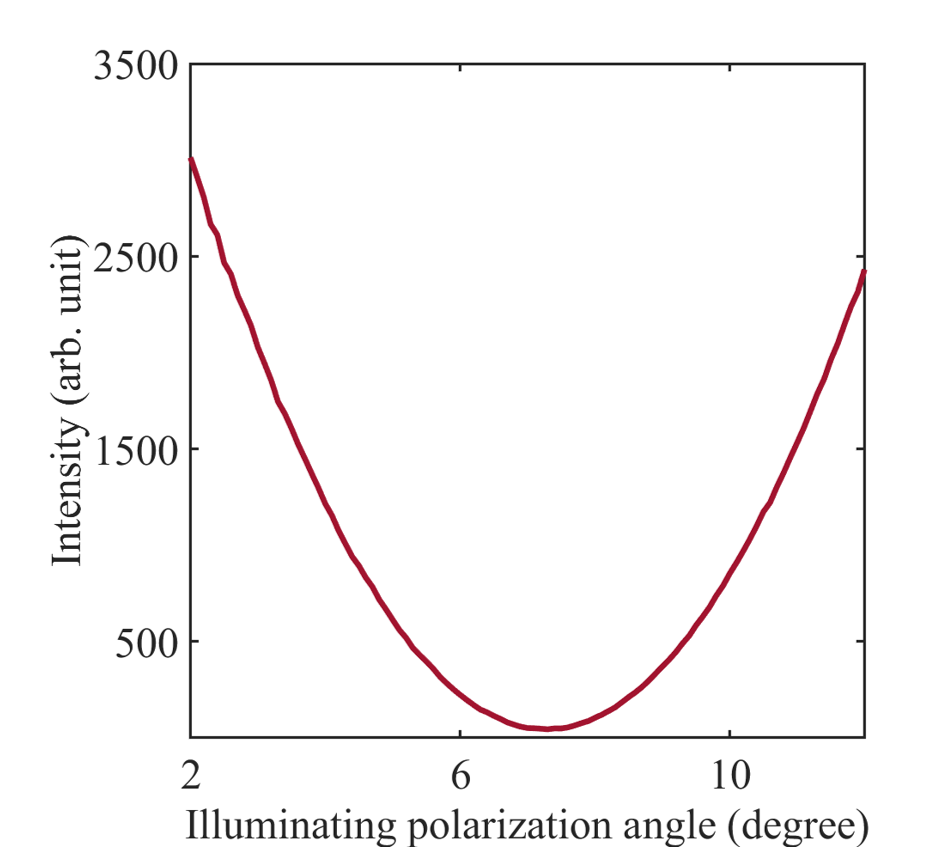}
\caption{Brown solid line illustrates the variation of the intensity of the output beam passing therough the tilted polarizer with changing angle of the illuminating polarization at $\phi=30^{o}$ and  $\psi=0^{o}$. The illuminating polarization angle noted are not the angle of polarization in the laboratory frame, but are the angle of the rotational mount attached with P.}
\label{figs1}
\end{figure}
 Eq.\eqref{eqs1} suggests that the illuminating polarization $\ket{E}$ should nearly coincide with $\hat{\parallel}$ to achieve maximum shift from the tilted polarizer. $\hat{\parallel}$ is the direction of the absorption axis of the polarizer. Hence, the large shifts occur when the output beam intensity is very low. Similar situation also arises for Brewster's angle scenario in transverse beam shift from a partial reflection. Fig.\ref{figs1} demonstrates the variation of the intensity with changing the angle of the illuminating polarization for tilt angle $\phi=30^{o}$. From Fig.2(a) of the main text and Fig \ref{figs1}, it is clear that the maximum shift occurs around cross polarization incidence.

\subsection*{S3. Experimental parameters}
\label{s3}
\begin{figure}[h!]
\centering
\includegraphics[width=0.3\linewidth]{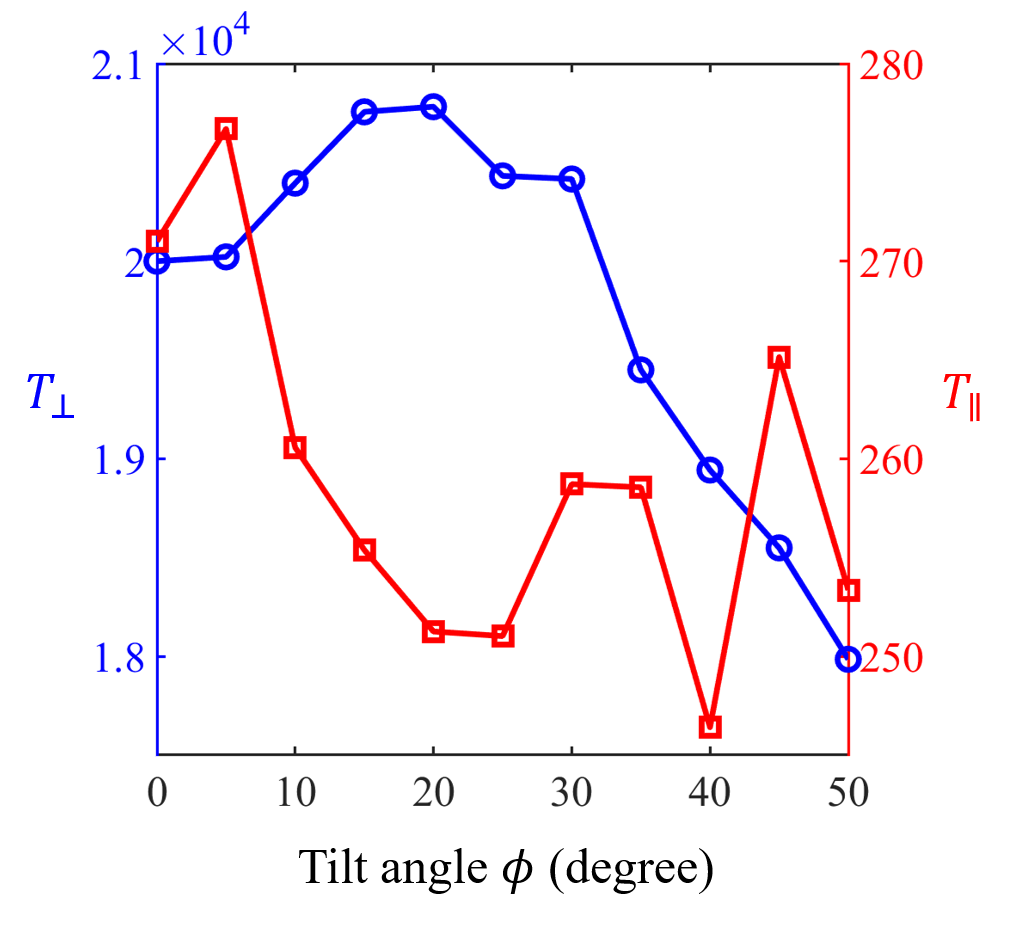}
\caption{The variation of $T_{\perp}$ (blue circled line) and $T_{\parallel}$ (red squared line) with changing tilt angle $\phi$ at azimuthal orientation $\psi=0$.}
\label{figs2}
\end{figure}

We have experimentally measured the transmission coefficients $T_{\parallel}$ and $T_{\perp}$ with changing tilt angle $\phi$ at a fixed azimuthal angle $\psi=0$. Fig.\ref{figs2} demonstrates the same. Fig.\ref{figs2} indicates that for our experimental regime, $\frac{T_\perp}{T_\parallel}\sim10^2\gg 1$. This approximation is also supported by the previous reports \cite{bliokh2019spin,korger2014observation}. It is reasonable to take $T_{\parallel}$ and $T_{\perp}$ as constant over the experimental regime. We take $T_{\parallel}\equiv<T_{\parallel}>\approx258.92$ and $T_{\perp}\equiv<T_{\perp}>\approx1.98\times10^4$.
\par
In our experimental set up, the parameters mentioned Eq.(4) and (6) of the maintext, take the following values. The wavelength of the input beam $=\frac{2\pi}{k}=633\ nm$, the Rayleigh range $Z_0=1.37\ mm$, the propagation distance after passing SP is $Z=18\ cm$. As mentioned, we take $T_{\parallel}\equiv<T_{\parallel}>\approx258.92$ and $T_{\perp}\equiv<T_{\perp}>\approx1.98\times10^4$.
\clearpage
\newpage
\bibliographystyle{ieeetr}
\bibliography{ref}

\end{document}